\documentclass{article}

\usepackage{arxiv}

\usepackage[utf8]{inputenc} 
\usepackage[T1]{fontenc}    
\usepackage{hyperref}       
\usepackage{url}            
\usepackage{booktabs}       
\usepackage{amsfonts}       
\usepackage{nicefrac}       
\usepackage{microtype}      
\usepackage{lipsum}
\usepackage{fancyhdr}       
\usepackage{graphicx}       
\graphicspath{{media/}}     
\usepackage{natbib}

\title{Does rainfall create buoyant forcing at the ocean surface?
}

\author{
  Dipanjan Chaudhuri$^*$, Eric D'Asaro \\
Applied Physics Laboratory\\
University of Washington\\
Seattle\\
\texttt{$^*$dipabkp@uw.edu} \\
}

\begin{document}
\maketitle

\begin{abstract}
Rain affects the buoyancy of the upper ocean in two ways: The freshwater flux in rain makes the water fresher and lighter, stabilizing the ocean (a negative buoyancy flux). The convective systems that produce rain are often accompanied by cold, dry air, often called `cold pools', and reduced short-wave radiation, which makes the water colder and heavier, destabilizing the ocean (a positive buoyancy flux). We estimate net buoyancy fluxes using \textit{in situ} measurements from twenty-two moored buoys in the equatorial oceans under different rainfall categories. We find that buoyancy fluxes tend to destabilize the ocean during light rain (0.2-4 mm/hr) and stabilize the ocean during heavy rain ($>$4 mm/hr). Furthermore, buoyancy fluxes during rain tend to be more positive at night than during the day, with nighttime rain twice as likely to cause instability compared to daytime rain, even at the same rainfall intensity. Average buoyancy fluxes across the tropics during rain can have either sign. These findings challenge the common assumption that rainfall makes the ocean surface lighter and provide a starting point for focusing on the overall effect of precipitation on the ocean.
\end{abstract}


\section{Introduction}\label{sec1}

The density of seawater is essential for global ocean circulation. If the density of a parcel of water is higher than the surrounding water, then it will sink due to gravitational instability. Thus, the lightest water stays at the surface in equilibrium, whereas the densest water settles at the bottom of the ocean. The density of seawater, which is determined from the Equation of state \cite{millero2010,cushman2011}, is primarily a nonlinear function of temperature, salinity, and pressure.  At the ocean surface, the linearized Equation of state is: 
\begin{equation}
\label{eq:BUOY_01}
\rho = \rho_{0}[1-\alpha(T-T_{0})+\beta(S-S_{0})]  
\end{equation}
where $\rho_{0}$, $T_{0}$, and $S_{0}$ are reference values of density, temperature and salinity of the sea water. $\alpha = \rho^{-1}\frac{\partial \rho}{\partial T}$ is the effective coefficient of thermal expansion, and $\beta = \rho^{-1}\frac{\partial \rho}{\partial S}$ is the coefficient of haline contraction.  Thus, salinity and temperature have an opposing effect on density. Thus, warmer and fresher surface waters in the tropics are lighter than colder and saltier surface waters in the subtropical oceans \cite{talley2011}.

The air-sea fluxes of heat (heating and cooling) and freshwater (evaporation, precipitation) modify the density of seawater. Buoyancy flux ($B_{0}$; m\textsuperscript{2}/s\textsuperscript{3}), which is composed of net surface heat flux ($Q_N$; W/m\textsuperscript{2}), rate of evaporation ($E$; m/s), and precipitation ($P$; m/s) is expressed as:
\begin{equation}
\label{eq:BUOY_02}
B_{0} = -\frac{g \alpha Q_N}{\rho C_P}+g\beta(E-P)S_{0}
\end{equation}     
where $g$ (m/s\textsuperscript{2}) is gravity, $\rho$ (kg/m\textsuperscript{3}) is ocean density, $C_P$ (J/kg/K) is specific heat of water, and $S_0$ is surface salinity. One can expect a more buoyant ocean surface in the presence of negative buoyancy flux due to either surface warming or precipitation. Note that in terms of buoyancy flux, a heat flux of 10 W/m\textsuperscript{2} into the ocean is approximately equivalent to a freshwater flux of 3 mm/day for an ocean with sea surface temperature (SST) of 29$^\circ$C and sea surface salinity of 32.5 psu.

The global ocean circulation is primarily driven by the interaction between buoyancy forcing at the air-sea interface and surface wind stress \citep{munk1950wind, Anderson1996,wunsch2004vertical,stewart2008,HOGG2010,HOGG2020}. Previous studies have shown that the difference in buoyancy fluxes at the ocean surface, between the tropics (20$^\circ$S to 20$^\circ$N) and the polar region, significantly impacts deep overturning circulations \citep{rahmstorf2003thermohaline,karnauskas2020physical}. This buoyancy gain in the tropics results from net heat gain, abundant rainfall in the Intertropical Convergence Zone (ITCZ, \cite{barry2009atmosphere}), and the proximity to major river mouths such as the Amazon, the Congo, the Ganges, and others \citep{DAI2002,Fekete2002}.

In oceanography, it is commonly assumed that rainfall always adds buoyancy to the ocean surface \citep{talley2011, Jacob2007, Neetu2012,balaguru2012}. However, tropical rain is usually followed by intense heat loss associated with atmospheric cold pools. These cold pools with a lifespan of up to 12 hours or more are formed by the evaporative cooling of falling rain, followed by a strong downward flow of cold and dry air within clusters of tall convective clouds in the tropics \citep{Zipser1977,houze2014, van2021}. When these reach the surface, they spread outward across the ocean surface as fast gravity currents over diameters ranging from 10 to 200 kilometers, making them several times larger than their associated rainfall areas of approximately 5 to 20 kilometers \citep{zuidema2017,deszoeke2017}. Measurements show that these significantly impact the air-sea fluxes \citep{WILLS2021,joseph2021, IURY2023}, leading to cooling and increased salinity in the ocean due to wind-induced and convective mixing \citep{Walesby2015, Reverdin2020, Iyer2021}, producing effects opposite to rainfall. The net buoyancy flux results from a balance between the positive buoyancy flux of rain and the negative flux of the accompanying cold pools.  Here, we investigate buoyancy flux during rainfall in the tropics using all available air-sea flux and rain measurements from tropical ($\pm20^{\circ}$) moorings. We aim to answer the simple question: ``Does rainfall generate a buoyant force on the ocean surface?"

\begin{figure}[!]
\centering
	\noindent\includegraphics[width=37pc,angle=0]{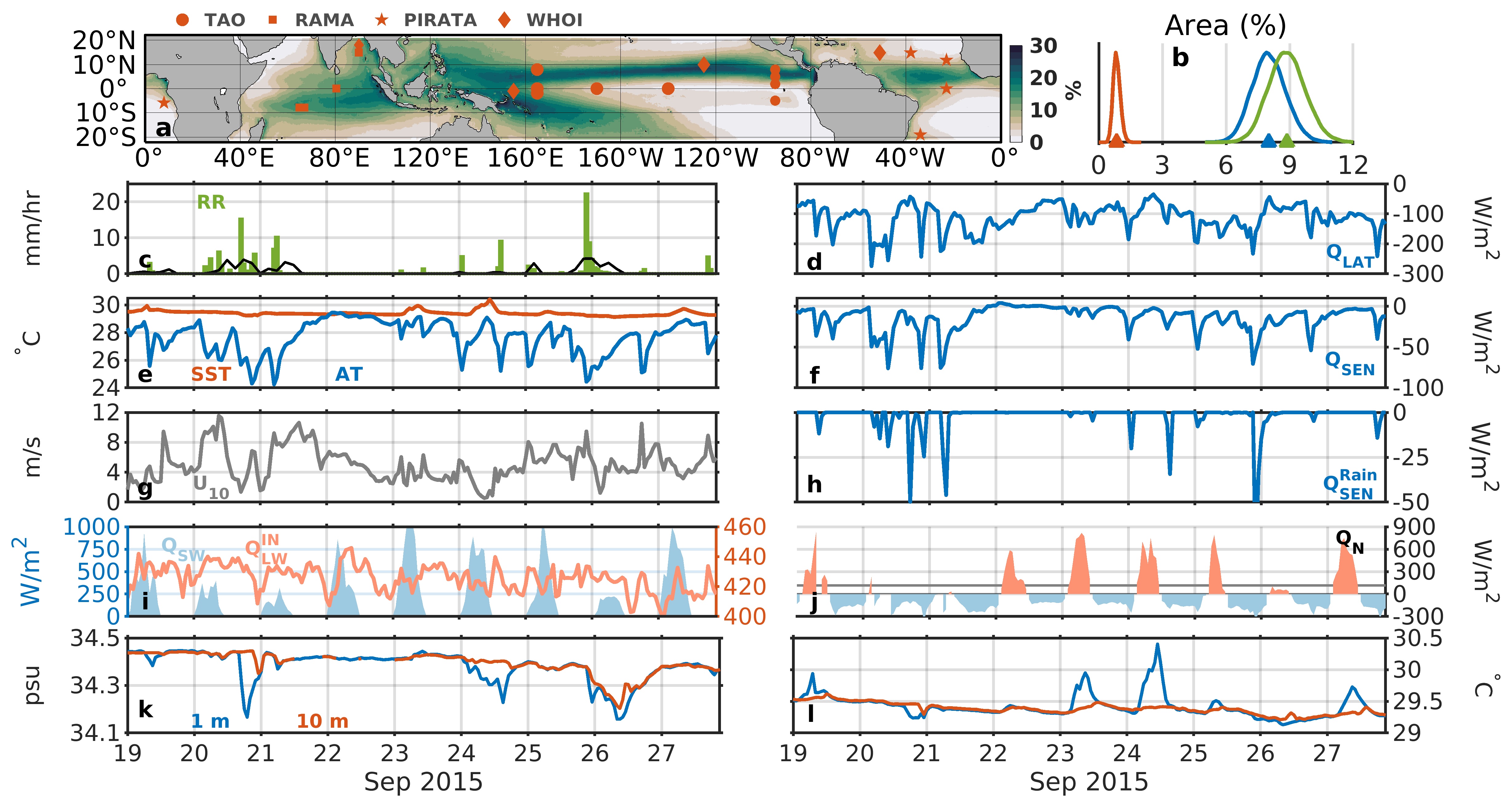}\\
	\caption{(a) Map showing the annual mean of the precipitation occurrence time ($\%$; color) and the mooring locations used in this study. The red-filled circles, squares, stars, and diamonds represent the TAO, RAMA, PIRATA, and WHOI moorings array. (b) PDFs of ocean-only fractions of the area under total (green), heavy (red), and light (blue) rain at any time. Upward triangles represent medians. The above estimations of annual mean of the precipitation occurrence time and ocean-only fractions of the area under rain at any time are based on twelve years (2008--2019) of the multi-satellite TRMM 3B42 3-hourly dataset and above a 0.2 mm/hr threshold over the equatorial ocean between 20$^\circ$S to 20$^\circ$N. Sample time series of hourly observations from the equatorial RAMA mooring at 0$^\circ$N 80.5$^\circ$E. (c) Rain rate ($RR$, green, mm/hr), (e) temperature at 1 m depth ($SST$, red, $^\circ$C), and air temperature ($AT$, blue, $^\circ$C), (g) wind speed ($U_{10}$, gray, m/s), (i) incident shortwave ($Q_{SW}$, light blue, W/m\textsuperscript{2}) and incoming longwave ($Q_{LW}^{IN}$, light red, W/m\textsuperscript{2}), (d) latent heat flux ($Q_{LAT}$, blue, W/m\textsuperscript{2}), (f) sensible heat flux ($Q_{SEN}$, blue, W/m\textsuperscript{2}), (h) rain-induced sensible heat flux ($Q_{SEN}^{Rain}$, blue, W/m\textsuperscript{2}), and (j) net heat flux ($Q_N$, W/m\textsuperscript{2}). Light red and light blue shaded areas in panel (j) represent positive (heat gain) and negative (heat loss) heat flux. The horizontal line in panel (j) marks the mean net heat gain during a cloud-free day. (k) Salinity and (l) temperature at 1 m (blue) and 10 m (red) depths.}\label{f1}
\end{figure}
\section{Results}\label{sec2}

\subsection{Background}\label{subsec2}

Before discussing the net buoyancy flux during rainfall, we examine the frequency of rain over the tropics. Based on reliable satellite-derived rainfall data, we estimate that precipitation exceeding 0.2 mm/hr occurs on average over 8\% of the tropical ocean in a year (see Figure \ref{f1}a). The highest frequency of rainfall is observed in the equatorial Pacific region. Furthermore, Figure \ref{f1}b demonstrates that at any given moment, rainfall occurs over an average of 9\% of the total tropical ocean area, which is roughly equivalent to the size of Canada. Approximately 1\% of the area experiences heavy rainfall (exceeding 4 mm/hr), while 8\% of the area experiences light rainfall (0.2--4 mm/hr). These findings are consistent with previous research, indicating that tropical rain over the ocean varies significantly across different time and spatial scales \citep{Teo2017, Trenberth2018}.

To set the context, we begin with a nine-day time series of rain rate (RR), sea surface temperature (SST), atmospheric temperature (AT), wind speed ($U_{10}$), and all heat fluxes at the equatorial RAMA mooring (0$^\circ$N 80.5$^\circ$E). Figure \ref{f1}c shows two clusters of rainfall events associated with the westward-propagating convective disturbances observed over the western Pacific and Indian Oceans \citep{Zuluaga2013,Yu2018}. Throughout the period, there are rain showers that last for about two to four hours every six hours, except on September 20, 2015, when it rains continuously for twenty hours. Compared to air temperature, the SST changes are modest after the rain, while the air temperatures instantly drop by 3--4$^\circ$C (Figure \ref{f1}e). These are called atmospheric ``cold pools," which are widely occurring phenomena in the equatorial region (see introduction ). 
Note that changes in the surface wind speed follow the changes in atmospheric temperatures closely (Figure \ref{f1}g), indicating that cold air's horizontal spread along the surface can alter the mean wind speed during cold pool events.

There are six rainy days when tall clouds in the tropics blocked incoming shortwave radiation \citep{Anderson1996,WELLER1996,WELLER2019}, and a minimum of 200 W/m\textsuperscript{2} is observed at noon on September 26, 2015 (Figure \ref{f1}i). The observed incoming longwave radiation for nine days varies within 20 W/m\textsuperscript{2} from its mean of -420 W/m\textsuperscript{2} (Figure \ref{f1}i).

 On these overcast days, cool and dry downdrafts led to larger latent heat and sensible heat loss (see Figure \ref{f1}d, f). The largest latent flux during this period reaches close to -290 W/m\textsuperscript{2}, while the largest hourly sensible heat flux is about -80 W/m\textsuperscript{2}, both associated with a rain event early on September 20 \citep{Anderson1996,WELLER1996,WILLS2021,IURY2023}. The sensible heat flux due to rain is smaller than the latent heat flux, but comparable (two-thirds) with the sensible heat loss plotted in Figure \ref{f1}h \citep{GOSNELL1995, FLAMENT1995,ANDERSON1998}. 
 
Thus, during rain events, a negative net heat flux persists at the sea surface due to the combined effect of these four factors (see Figure \ref{f1}j), contributing to the net heat loss of -120 W/m\textsuperscript{2} for two consecutive days (from September 20 to 21) and for one day on September 26. While these observations are from one mooring, heat losses at the sea surface during rain are common \citep{Anderson1996,WELLER1996,ANDERSON1998,WELLER2019,IURY2023}. More figures that show exact quantities, as in Figure \ref{f1}c-j from the moored observation at different locations of the tropical oceans, are provided in the supporting information (see Figures 1--4 in the supporting information).

\begin{figure}[!]
\centering
	\noindent\includegraphics[width=31pc,angle=0]{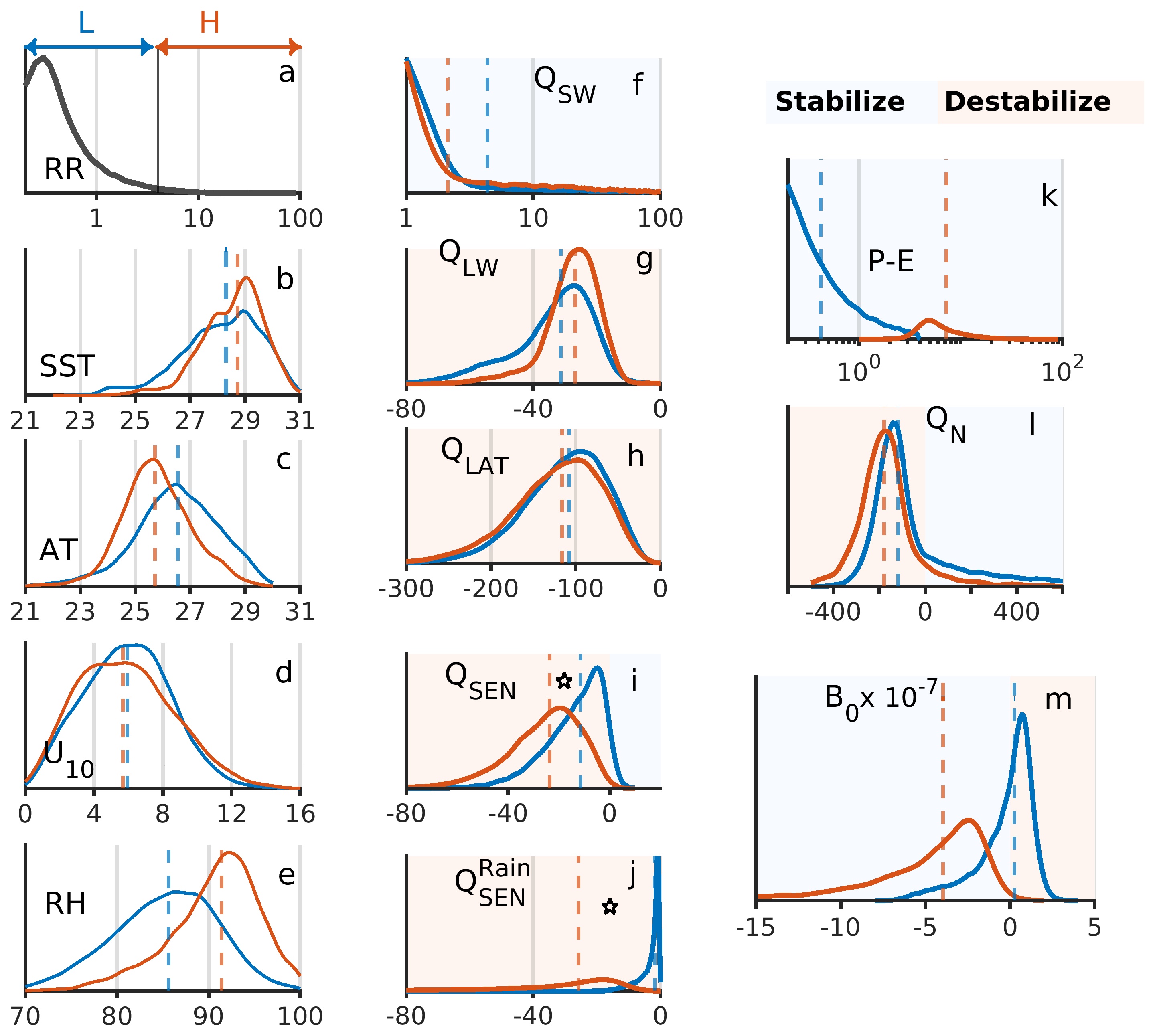}\\
	\caption{Probability density function (PDF) of (a) rain rate ($RR$, grey, mm/hr), (b) temperature at 1 m depth (SST, $^\circ$C), (c) air temperature (AT, $^\circ$C), (d) wind speed ($U_{10}$, m/s), (e) relative humidity ($RH$, \%), (f) shortwave radiation ($Q_{SW}$, W/m\textsuperscript{2}), (g) net longwave radiative heat flux ($Q_{LW}$, W/m\textsuperscript{2}), (h) latent heat flux ($Q_{LAT}$, W/m\textsuperscript{2}), (i) sensible heat flux ($Q_{SEN}$, W/m\textsuperscript{2}), (j) rain-induced sensible heat flux ($Q_{SEN}^{Rain}$, W/m\textsuperscript{2}), (k) precipitation minus evaporation ($P-E$, mm/hr), (l) net heat flux ($Q_N$, W/m\textsuperscript{2}), and (m) buoyancy flux ($B_{0}$, m\textsuperscript{2}/s\textsuperscript{3}) for `Light' (L, blue) and `heavy' (H, red) rain. The vertical line (black) in panel (a) marks the rain rate 4 mm/hr. The dashed vertical lines in the panels (b-m) represent the median of the probability density functions. Light red and light blue colored shading in the panels (f-m) represent destabilizing and stabilizing forcing regimes. The medians for each variable are listed in Table 1 in the supporting information. In panels i and j, a star highlights the difference between the two components of sensible heat fluxes under heavy and light rain.}\label{f2}
\end{figure}

During rain events, we observe the formation of four near-surface freshwater anomalies, commonly referred to as rain puddles. These anomalies have salinity stratifications ranging from 0.05 to 0.2 psu within the upper 10 meters of the ocean, while the temperature anomalies remain relatively modest (Figure\ref{f1}k, l). These rain puddles form under very low wind conditions, probably occurring when the surface divergence of cold pools opposes the mean wind flow (Figure\ref{f1}g). Two rain puddles are associated with significant surface heat loss linked to strong cold pools on September 20 and 26, while the other two emerge during heat gain on September 19 and 23, when the cold pools are weaker (Figure\ref{f1}j).

A notable feature of these rain puddles is that a greater stratification of vertical salinity is observed in the upper ocean during heat gain phases, although these events experience less rainfall compared to those associated with heat loss, which can trigger convective mixing (Figure\ref{f1}j). These observations highlight the important role of thermal buoyancy fluxes during rain over the ocean in setting up upper ocean vertical stratification, suggesting the need for further investigation.

\subsubsection{Rainfall Intensity}\label{subsubsec2}
We now investigate the effect of rain intensity on the thermal and haline buoyancy fluxes of light (0.2--4 mm/hr) and heavy (exceeding 4 mm/hr) rainfall across tropical oceans. Light rain is more frequent than heavy rain, occurring 84\% of the time (Figure \ref{f2}a). Heavy rain is more likely to occur over warmer sea surface temperatures (SST; Figure \ref{f2}b). It is also associated with cooler air temperatures (Figure \ref{f2}c), which suggests a correlation between stronger cold pools and heavy rain. The SST is consistently warmer than the air temperature, with a median difference of 2.8$^\circ$C, which accounts for more sensible cooling. The observed wind speed ($U_{10}$) ranges from 0 to 16 m/s, with a median of 6 m/s for both rainfall categories (Figure \ref{f2}d). Note that the larger spread of $U_{10}$ for heavy rainfall likely results from the stronger outflows associated with cold pools, which may either enhance or weaken the mean wind (Figure \ref{f2}d).

The median values of $Q_{SW}$ for light and heavy rain are 5 W/m\textsuperscript{2} and 2 W/m\textsuperscript{2}, indicating that the cloud blocks incoming sunlight (Figure \ref{f2}f). Clouds can also reduce the net longwave radiation ($Q_{LW}$) by trapping the outgoing longwave. The probability distribution functions (PDFs) of $Q_{LW}$ during light rain show a higher prevalence of negative values due to the absence of tall convective clouds (Figure \ref{f2}g). 

The largest cooling occurs at the ocean surface during rain due to the latent heat flux ($Q_{LAT}$). Interestingly, the PDFs of $Q_{LAT}$ for both categories are similar (Figure \ref{f2}h), with median values of -107 W/m\textsuperscript{2} and -116 W/m\textsuperscript{2} for light and heavy rainfall, respectively. This similarity can be attributed to comparable humidity differences ($q _s - q_a$) between the surface and the atmosphere, which remain consistent across similar wind speeds (Figure \ref{f2}d) for both rainfall categories. Heavy rain, which is characterized by warmer SST, colder air temperature,  and higher atmospheric RH (median value of 91\%), has similar effects on the specific humidity gradient to light rain, which occurs in slightly lower SST, warmer air temperature, and lower RH (median value of 85\%) according to the Clausius Clapeyron equation (Figure \ref{f2}b, c, e, and Table 1 in the supporting information)). These factors lead to similar evaporative cooling under different rain conditions.   

The variability of the sensible heat flux ($Q_{SEN}$) mirrors the differences in air-sea temperatures, with about 50\% more cooling for heavy rainfall than light rainfall (Figure \ref{f2}i). The median values of the sensible heat flux due to rain ($Q_{SEN}^{Rain}$) are -2 W/m\textsuperscript{2} and -25 W/m\textsuperscript{2} for light and heavy rainfall, respectively (Figure \ref{f2}j). Note that the sensible heat due to rain (-24 W/m\textsuperscript{2} difference) causes more cooling than the normal sensible heat flux (-12 W/m\textsuperscript{2}) during heavy rain than in light rain (see Table 1 in the supporting information).

As expected, the impact of evaporation is more pronounced in light rain than in heavy rain; precipitation minus evaporation ($P-E$) medians are reduced by 30\% and 3\% for light and heavy rain (Figure \ref{f2}k). Total heat fluxes during rain have negative and positive values (mainly during the daytime when $Q_{SW}$ dominates the total heat fluxes). Figure \ref{f2}l shows that $Q_N$ associated with heavy rainfall is colder (-180 W/m\textsuperscript{2}) than $Q_N$ under light rainfall (-120 W/m\textsuperscript{2}), primarily due to larger sensible heat fluxes (Figure \ref{f2}i, j), as other components of the heat fluxes are comparable for both types of rain. Nevertheless, the net buoyancy fluxes ($B_0$) under heavy rainfall remain negative over the equatorial ocean. Meanwhile, 60\% of the time, $B_0$ associated with light rain is positive, meaning that thermal buoyancy fluxes overcome the stabilizing effect of haline buoyancy fluxes (as shown in Figure \ref{f2}m). Therefore, heavy rain typically has a stabilizing net buoyancy force, whereas light rain is often accompanied by destabilizing buoyancy forces at the ocean surface.

 \begin{figure}[t!]
 \centering

	\noindent\includegraphics[width=17pc,angle=0]{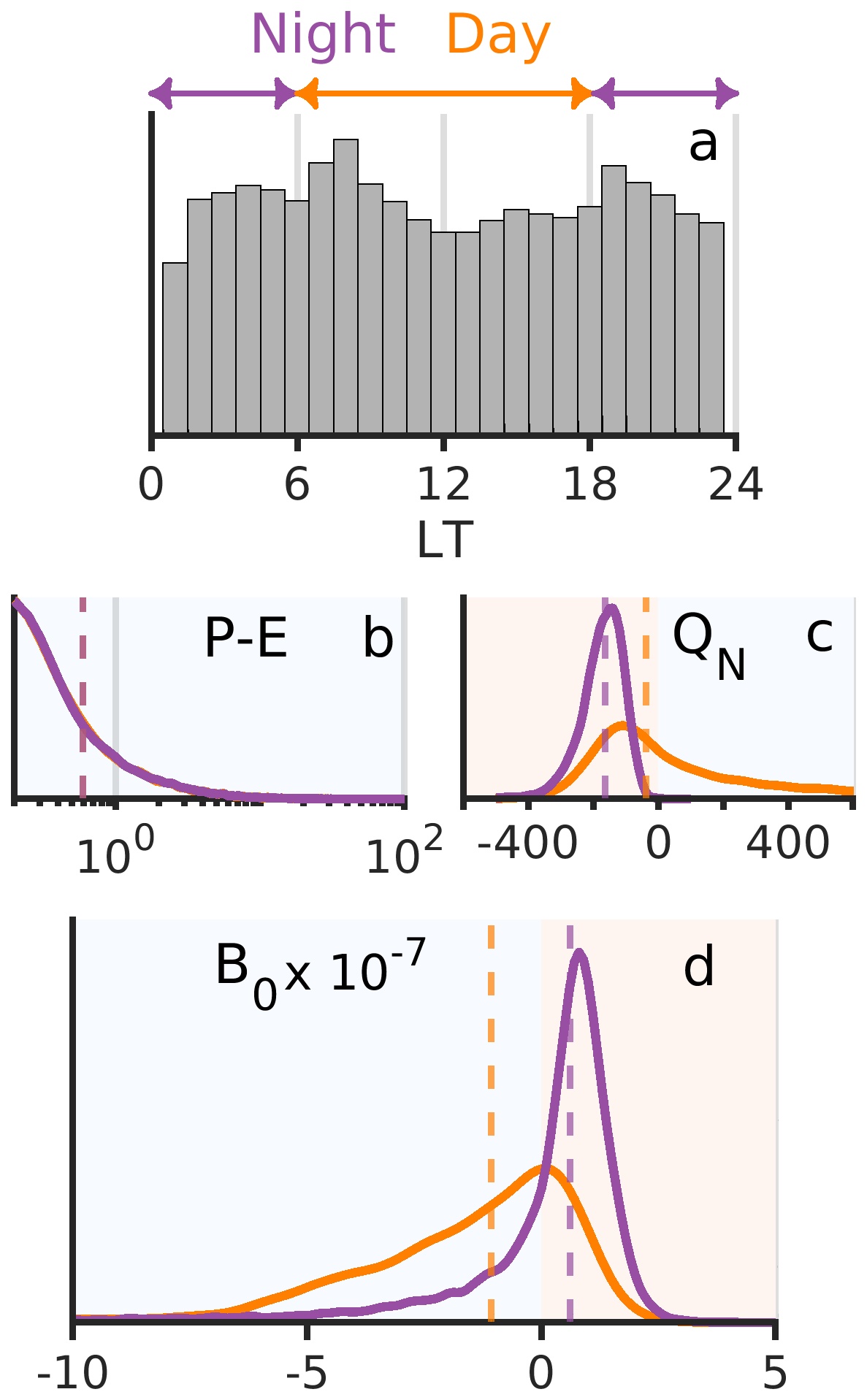}\\
	\caption{(a) Normalized histograms of diurnal variation of rainfall events across the tropical ocean. Probability density function (PDF) of (b) precipitation minus evaporation ($P-E$, mm/hr), (c) net heat flux ($Q_N$, W/m\textsuperscript{2}), and (d) buoyancy flux ($B_{0}$, m\textsuperscript{2}/s\textsuperscript{3}) for `day' (orange) and `night' (maroon) rain. The dashed vertical lines in the panels (b--d) represent the median of the probability density functions. Light red and light blue colored shading in the panels (b--d) represent destabilizing and stabilizing forcing regimes. The medians for each variable are listed in Table 2 in the supporting information.}\label{f3}
\end{figure}

\subsubsection{Diurnal variation}\label{subsubsec2a}

The diurnal cycle significantly influences tropical rainfall variability, primarily driven by nocturnal radiative cooling, sea surface temperature modulation, and atmospheric tides \citep{mapes1992,kikuchi2008,minobe2020,fang2022global}. This leads us to explore how buoyancy fluxes associated with rainfall differ between daytime (local time 6:00 AM to 6:00 PM) and nighttime (local time 12:00 AM to 6:00 AM and 6:00 PM to 12:00 AM) in the tropical ocean. Figure \ref{f3}a shows the rainfall pattern throughout the day, with two distinct peaks around 8:00 and 19:00 local time (LT), consistent with previous studies  \citep{liu2008,minobe2020}. Our analysis suggests that the chances of experiencing positive, that is, destabilizing $B_0$ are twice as high at night compared to day (Figure \ref{f3} c). Thus, night rain favors more active mixing and deeper boundary layers. The PDFs of $B_0$ exceed zero approximately 60\% and 30\% of the time during the night and day (Figure \ref{f3}d and Table 2 in the supporting information), respectively. This is due to the lack of $Q_{SW}$ at night, leading to greater heat losses than during the day (see Figure \ref{f3}b), even under the same haline fluxes (Figure \ref{f3}a). 

There are no significant differences in other components of heat fluxes, such as net longwave, latent, and sensible fluxes, nor in other meteorological variables such as sea surface temperature (SST), atmospheric temperature, and wind speed between day and night (see Figures 5--6 in the supporting information). As expected, during light rain at night, the thermal buoyancy fluxes overcome the stabilizing effect of the haline buoyancy fluxes 67\% most of the time. In contrast, this occurs only 18\% of the time during daytime light rain, mainly due to the presence of $Q_{SW}$ (as noted in the supporting information). For both heavy rain events, whether during the day or night, stable buoyancy fluxes are generated at the ocean's surface but with reduced intensity due to significant cooling (see the supporting information).

\subsubsection{Spatial variation}\label{subsubsec3}

\begin{figure}[!h]
\centering
	\noindent\includegraphics[width=33pc,angle=0]{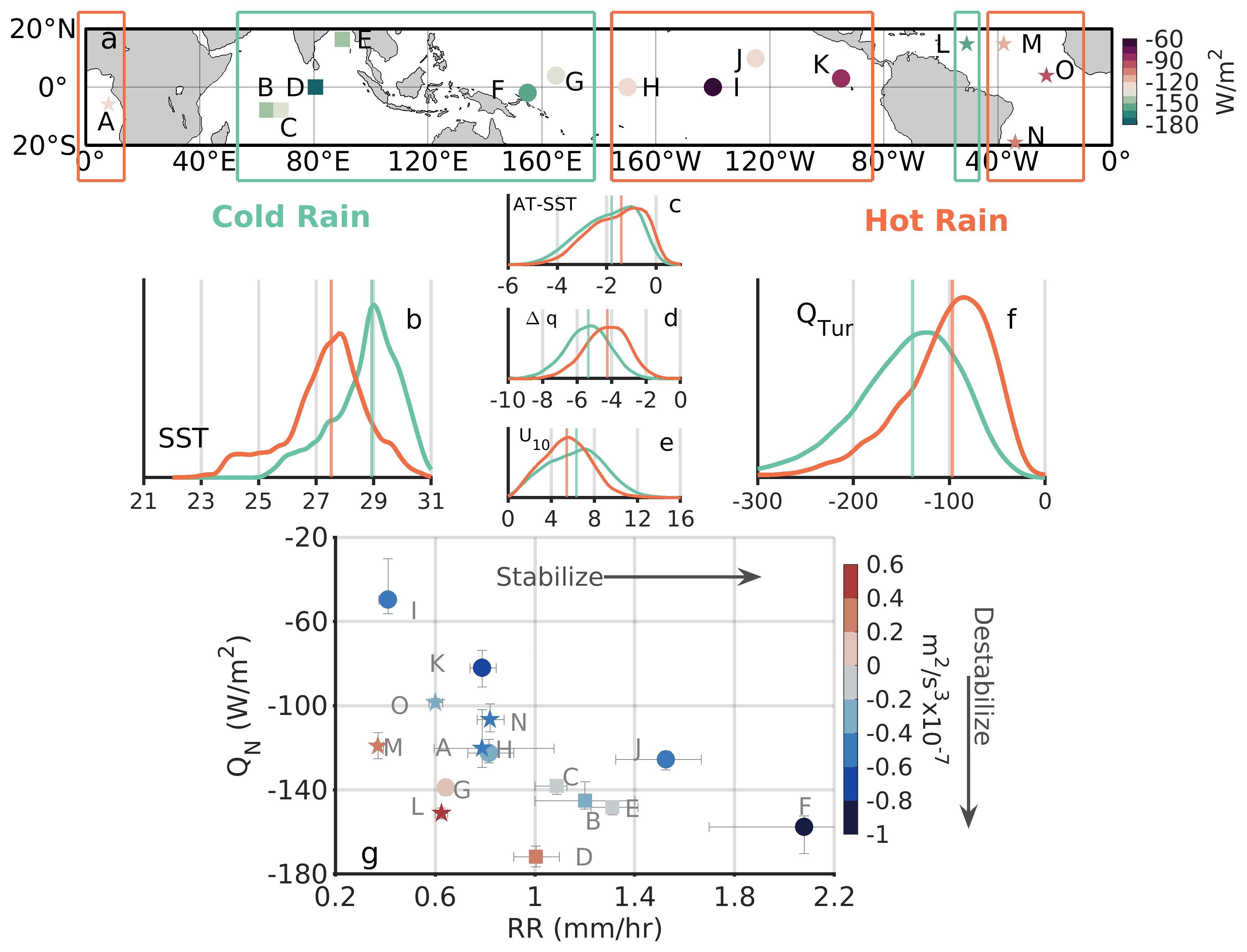}\\
	\caption{Geographical variation of heat fluxes during rainfall. (a) Stars, circles, and squares represent the position of moorings in the Atlantic, Indian, and Pacific oceans. Moorings at 8$^\circ$E, 65$^\circ$E, 67$^\circ$E, 80.5$^\circ$E, 90$^\circ$E, 155$^\circ$E, 165$^\circ$E, 170$^\circ$W, 140$^\circ$W, 125$^\circ$W, 95$^\circ$W, 51$^\circ$W, 38$^\circ$W, 34$^\circ$W, 23$^\circ$W are marked by \textbf{A}, \textbf{B}, \textbf{C}, \textbf{D}, \textbf{E}, \textbf{F}, \textbf{G}, \textbf{H}, \textbf{I}, \textbf{J}, \textbf{K}, \textbf{L}, \textbf{M}, \textbf{N}, and \textbf{O}. Color represents the median of the net net heat flux ($Q_N$, W/m\textsuperscript{2}). PDFs of (b) temperature at 1 m depth (SST, $^\circ$C), (c) temperature difference between air (AT, $^\circ$C) and sea surface (SST, $^\circ$C), (d) specific humidity difference between air and saturated specific
    humidity at SST ($\Delta q= q_a-q_s$; g/kg), (e) wind speed ($U_{10}$, m/s), and (f) turbulent heat flux ($Tu= Q_{SEN}+Q_{LAT}$, W/m\textsuperscript{2}) for Cold (light green) and Hot (light orange) rain. (g) Scatter plot of net buoyancy flux  ($B_0$, m\textsuperscript{2}/m\textsuperscript{3}) as a function of net heat flux ($Q_N$, W/m\textsuperscript{2}) and rain rate ($RR$, mm/hr). The color of the circles, stars, and squares represents the median of the net buoyancy flux measured at different moorings during rain events. Error bars are 95\% bootstrap estimates.}\label{f4}
\end{figure}
Rainfall intensity and the associated fluxes during rainfall vary spatially across the tropical oceans \citep{Zuluaga2013,biasutti2012very,fang2022global}. We analyzed a total of 150 buoy-years of data from twenty-two moorings, including 31000 hours of rain measurements throughout the equatorial region (see Methods and Table 3 in the supporting information for details). Figure \ref{f4}a illustrates the average net heat flux during rainfall, represented by colors for different longitudes; the moorings are grouped into fifteen longitude bins, with some groups, like Group \textbf{G}, containing multiple moorings. Note that the net heat flux at all locations is negative, indicating net heat loss from the ocean during rainfall. The averages shown are median values, with 95\% confidence limits determined through a bootstrap median analysis. We use medians instead of means because the data are skewed and contain outliers. Even if mean values were used, the overall conclusion remains valid qualitatively (see Figure 9 in the supporting information). Note that any of the two PDFs representing different scenarios, such as light vs. heavy, day vs. night, or cold vs. hot for various variables such as SST, heat fluxes, Buoyancy fluxes, etc, throughout the study, are statistically different from each other at a 5\% significance level, as indicated by the Kolmogorov-Smirnov test \citep{lilliefors1967}.

We identified two distinct geographical regions characterized by contrasting heat loss during tropical rain: ``Cold rain," which results in greater cooling, and ``Hot rain," which is linked to less cooling (Figure \ref{f4}a). Further analysis indicates that the differences are not due to the frequency of heavy versus light rain or the occurrence of more rain during nighttime, as both conditions usually contribute to increased cooling (see Figure 7 in the supporting information). Instead, we observe more heat losses in the form of sensible and latent heat fluxes from the ocean to the atmosphere (Figure \ref{f4}c, d, f);  the heat loss of -200 W/m\textsuperscript{2} in cold rain regions is double that of hot rain regions. These significant cooling events are linked to warm sea surface temperatures \citep{gadgil1984,graham1987,wallace1992}, which can trigger deep atmospheric convection followed by a strong downdraft (Figure \ref{f4}e) of colder air (Figure \ref{f4}c) that holds less water vapor (according to the Clausius-Clapeyron relation) during rainfall  (Figure \ref{f4}d). These findings indicate complex air-sea interactions and the potential for different cloud formations in two thermodynamically distinct environments \citep{Houze15a}, warranting further investigation.

We can represent the average buoyancy flux during rainfall (shown in colors) as a function of rainfall rate ($RR$) and net heat flux ($Q_N$) across different longitudes (Figure \ref{f4}g). The rainfall rate and heat flux are not correlated and cover a wide range of values. Thus, very negative buoyancy fluxes, i.e., freshwater dominated, can occur either for small negative heat fluxes and light rain, as at \textbf{I} in the central Pacific, or with large negative heat fluxes and heavy rain, as at \textbf{F} in the western Pacific. Similarly, large positive buoyancy fluxes, i.e., heat-dominated, occur with large negative heat fluxes and light rain, as at \textbf{L} in the central Atlantic.

\begin{figure}[!h]
\centering
	\noindent\includegraphics[width=15pc,angle=0]{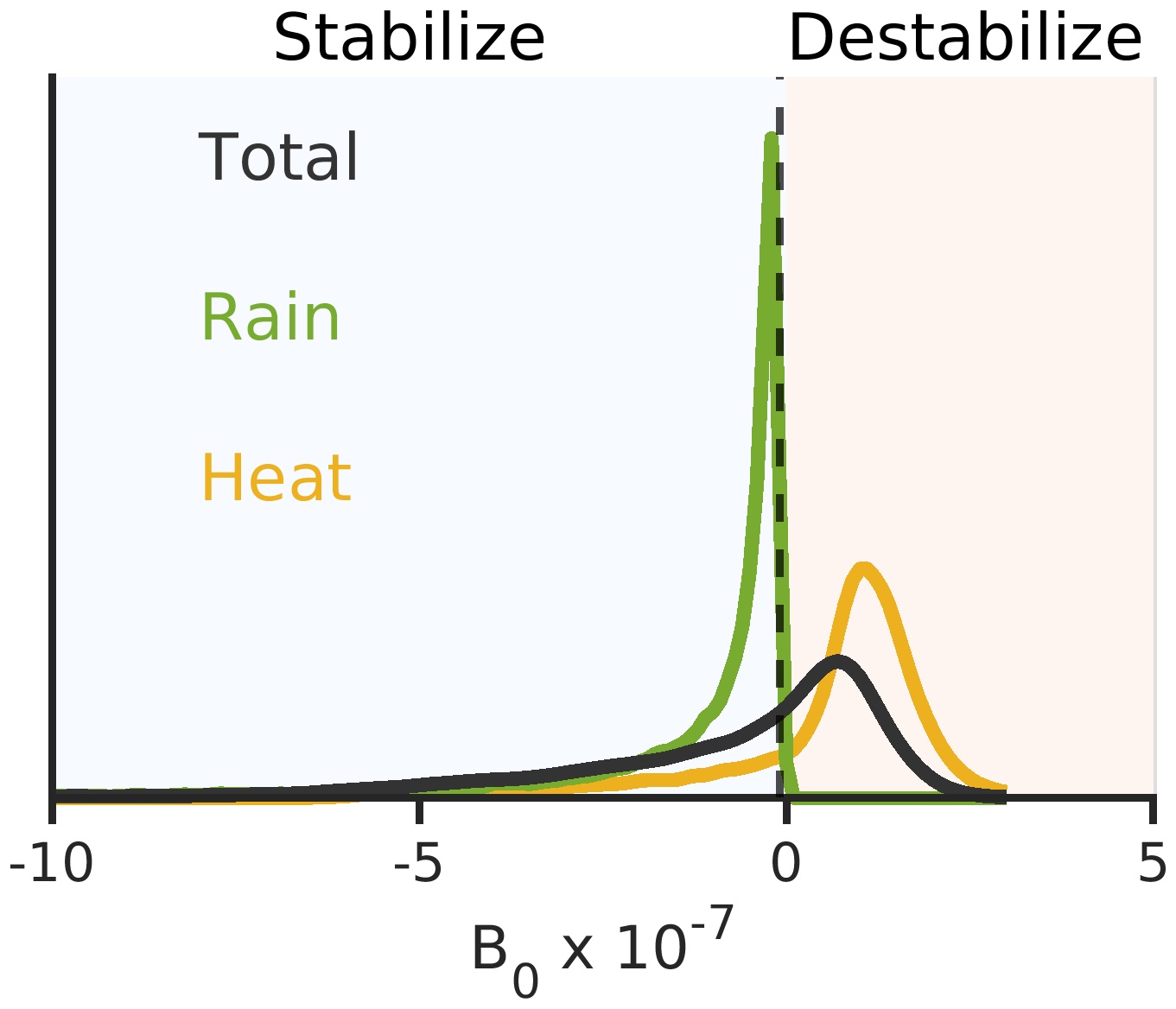}\\
	\caption{PDFs of net buoyancy flux (m\textsuperscript{2}/s\textsuperscript{3}; black), buoyancy flux due to rain (m\textsuperscript{2}/s\textsuperscript{3}; green), and  buoyancy flux due to heat (m\textsuperscript{2}/s\textsuperscript{3}; yellow).}\label{f5}
\end{figure}

\subsubsection{Net Buoyancy flux}\label{subsubsec3}

When we combine all the buoyancy fluxes, we find that the chance of getting positive (destabilizing) values is close to 50\% during rainfall events (Figure \ref{f5}). The stabilizing effect of rain (green) on the upper ocean is often offset by the a destabilizing effect of the associated cold pools and short-wave suppression (yellow). Considering only the former, without the latter, leads to large errors in the net buoyancy flux and thus the response of the upper ocean to rainfall. 

\section{Discussion}\label{sec12}

 Our study examines the buoyancy flux at the air-sea interface in the tropical ocean during rainfall using moored observations. We find:  
 
 \begin{itemize}
    \item Rainfall in the tropical ocean is intermittent, with an average of 8-9\% of the region experiencing precipitation annually or at any given moment. Light rain is more common than heavy rain, occurring 90\% of the time, consistent with previous findings.
    \item  During rain, there is a significant decrease in incoming shortwave radiation and an increase in turbulent air-sea fluxes (latent, sensible, and rain-induced sensible heat fluxes) due to atmospheric cold pools. This leads to a net heat loss. Sometimes, the heat losses can offset the stabilizing effect of rainfall and result in positive buoyancy fluxes.
    \item More heat loss occurs during heavy rain compared to light rain. This difference is mainly due to the cooling caused by the sensible heat carried by rain. Other components, including shortwave, longwave, and, surprisingly, latent heat fluxes, remain comparable for both types of rainfall. Heavy rain always leads to buoyant buoyancy fluxes, while the probability of destabilizing buoyancy fluxes under light rain is approximately 60\%. Evaporation related to cold pools is not a significant factor in the buoyancy budget, except during light rain, where it can reduce rain contribution by 30\%. 
    \item Our analysis suggests that the chances of experiencing positive, i.e., destabilizing, $B_0$ are twice as high during nighttime (60\%) rainfall compared to daytime (30\%) rainfall. This is due to the lack of $Q_{SW}$ at night, which leads to greater heat losses than during the daytime, even under the same haline fluxes.
     \item Our analysis identifies two regions: ``Cold rain," with more heat loss observed in the equatorial Indian Ocean and the west Pacific warm pool, and ``Hot rain," which experiences less cooling observed during rain in the central Pacific Ocean. The cooling differences are linked to frequent events that have increased sensible and latent heat losses in cold rain areas, not due to the occurrence of more light or night rain. These events are related to warm sea surface temperatures that trigger intense atmospheric convection and swift downdrafts of cooler, less humid air. 
    \item Overall, the chance of getting positive (destabilizing) buoyancy flux is close to 50\% during rainfall events. This challenges the usual belief that rain always causes positive buoyancy in the tropical ocean.
 
 \end{itemize}
 
 This study emphasizes that rain does not occur in isolation, but as part of convective atmospheric systems that dramatically change the air-sea fluxes in multiple complex ways depending on the type and intensity of the systems.  It suggests a focus on the overall effect of these systems on the ocean, on both large and small scales, rather than on just the rainfall. Immediate future questions can be asked: (a) What is the distribution of buoyancy fluxes associated with rain over the extratropical ocean? (b) How well do atmospheric reanalysis products capture buoyancy fluxes during rain? (c) Do the present climate models have the capacity to reproduce the observed rain buoyancy fluxes since changing intensity, frequency, and precipitation amounts are inevitable due to human-induced climate change \citep{Trenberth2003,Trenberth2011,bjorn2013}?

 \section{Methods}\label{sec13}
\subsection{Data}
In this study, we use hourly data from twenty-two different moored buoys (Figure \ref{f1}a) spanned over equatorial oceans (20$^\circ$S--20$^\circ$N): Five in the Indian Ocean, twelve in the Pacific Ocean, and six in the Atlantic Ocean. Fourteen of them are part of the Global Tropical Moored Buoy Array (GTMBA), which is a multi-national effort to improve both knowledge and understanding of the major ocean-atmosphere phenomena such as the El Ni\~{n}o-Southern Oscillation (ENSO), the Meridional Atlantic mode, and the Monsoons. These arrays of moored buoys are called the Research Moored Array for African-Asian-Australian Monsoon Analysis and Prediction \cite[RAMA]{RAMA2009} in the Indian Ocean, the Tropical Atmosphere Ocean \cite[TAO]{TAO1998} in the Pacific Ocean, and the Prediction and Research Moored Array in the Tropical Atlantic \cite[PIRATA]{PIRATA2008} in the Atlantic Ocean. Four moorings deployed by the Upper Ocean Processes Group (UOP) of the Woods Hole Oceanographic Institution (WHOI) in the equatorial ocean are used in this study \citep{COARE1996, BOB2016, SPURS2019}. We utilize a total of 150 buoy-years of data, with each buoy providing measurements spanning between 0.4 and 16.4 years. Our analysis focuses on 31000 hours of rain events recorded over all the available moored observations in the equatorial ocean. For further details, see Table 1 and Figure 8 in the supporting information. All twenty-two moorings are equipped with meteorological and oceanographic sensors. Bulk air-sea fluxes, e.g, latent and sensible heat flux, net heat flux, and other associated fluxes including evaporation, and evaporation minus precipitation are estimated using the COARE 3.0b algorithm \citep{Fairall2003}.

\subsection{Sensible heat flux due to rain ($Q_{SEN}^{Rain}$)}
here are two key components of ``sensible" heat exchange at the air-sea interface during rainfall: (i) the turbulent transfer of heat from the ocean surface to the atmosphere, referred to as sensible heat flux ($Q_{SEN}$), and (ii) the cooling of the ocean surface that occurs when cold raindrops fall onto the ocean and exchange heat with the surface, which is referred to as sensible heat flux due to rain ($Q_{SEN}^{Rain}$). The former, i.e., $Q_{SEN}$, is estimated at the ocean surface using a bulk formula according to the  COARE 3.0b algorithm \citep{Fairall2003}. The latter, i.e., $Q_{SEN}^{Rain}$, is calculated using the equation following \cite{FLAMENT1995}:

\begin{equation}
\label{eq:BUOY_03}
Q_{SEN}^{Rain}=\rho_wC_pRR(T_R-T_S)
\end{equation}  
where $\rho_w$ is the density of water, $RR$ is the rain rate, $T_R$ is the temperature of the rain, $T_S$ is the temperature of the surface. The assumption is that the falling raindrops' temperature equates to the atmosphere's wet-bulb temperature ($T_w$). This equilibrium depends on both the size of the raindrop and its velocity through the air. Previous model and observational studies \citep{FLAMENT1995,GOSNELL1995,ANDERSON1998} have indicated that this wet-bulb assumption is reasonable, with an uncertainty of about $\pm$0.4 K. We estimate $Q_{SEN}^{Rain}$ using equation \ref{eq:BUOY_03} and the standard psychrometer equation $T_w=T_a-\frac{Ld_v}{d_hC_p}(q_s(T_w)-q)$ consistent with the method employed by \cite{COARE1996}. $L$ is the latent heat of vaporization of water, $T_a$ is the ambient temperature, $d_v$ and $d_h$ are the diffusivities for water vapor and heat, $q_s$ is the saturation humidity at the $T_w$, and $q$ is the specific humidity at the ambient temperature.

\subsection{Rain gauge error estimates}
In this study, all twenty-three moorings are equipped with R. M. Young self-siphoning rain gauges that collect rainfall data every minute during measurement periods. Previous studies \citep{serra2001,serra2004} indicated that the overall error in the derived rain rate, primarily due to undercatch caused by wind, is 1.3 mm/hr for 1-minute data and 0.4 mm/hr for 10-minute data. For hourly data, the instrument noise is reduced by a factor of $1/(N)^{1/2}$, resulting in an error of 0.16 mm/hr, where N = 6 represents the number of 10-minute samples averaged over an hour. Therefore, we chose a lower limit of 0.2 mm/hr for light rain in this study.
\subsection{Data availability}

The data analyzed in this article are all publicly available. Data from the WHOI moorings is available at \url{https://uop.whoi.edu/index.html}, and data from the PMEL moorings are available at \url{https://www.pmel.noaa.gov/tao/drupal/flux/index.html}. TRMM 3B42 rainfall data is freely available at \url{https://trmm.gsfc.nasa.gov/3b42.html}.

\section*{Acknowledgments}
This work is supported by ONR grant N00014-17-1-2728. D.C. thanks Prof. Debasis Sengupta for the helpful discussion. The authors thank the GTMBA of the PMEL and the UOP of the WHOI for providing the moored data.

\bibliographystyle{unsrt}  
\bibliography{BUOYANCY_RAINFALL.bib}

\end{document}